\documentclass[11pt,a4paper]{article}
\usepackage{jheppub,feynmp,tikz}
\usepackage{epstopdf}

\usepackage{amssymb}
\usepackage{graphics,bm}
\usepackage{epsfig}
\usepackage{graphicx}
\usepackage{rotating}
\usepackage{color}
\usepackage{ulem}

\newcommand{\beq}{\begin{equation}}
\newcommand{\eeq}{\end{equation}}
\newcommand{\bqa}{\begin{eqnarray}}
\newcommand{\eqa}{\end{eqnarray}}

\def\square{\vcenter{\vbox{\hrule height.4pt
          \hbox{\vrule width.4pt height4pt
          \kern4pt\vrule width.3pt}\hrule height.4pt}}}

\title{Inverse magnetic catalysis and regularization in the quark-meson model}

\author[a]{Jens O. Andersen}
\date{today}
\affiliation[a]{Department of Physics, 
Norwegian University of Science and Technology, 
H{\o}gskoleringen 5,
N-7491 Trondheim, Norway}
\emailAdd{andersen@tf.phys.ntnu.no}
\emailAdd{william.naylor@ntnu.no}
\author[a]{William R. Naylor}
\author[b]{Anders Tranberg}
\emailAdd{anders.tranberg@uis.no}
\affiliation[b]{
Faculty of Science and Technology, University of Stavanger,
N-4036 Stavanger, Norway}

\abstract{
Motivated by recent work on inverse magnetic catalysis at finite
temperature, we study the
quark-meson model using both dimensional regularization and a sharp cutoff.
We calculate the critical temperature for
the chiral transition as a function of the Yukawa coupling in the
mean-field approximation varying the renormalization scale 
and the value of the ultraviolet cutoff.
We show that the results depend sensitively on how one treats the fermionic
vacuum fluctuations in the model and in particular on the regulator
used.
Finally, we explore a $B$-dependent transition temperature for the
Polyakov loop potential $T_0(B)$ using the functional renormalization group.
These results show that even arbitrary freedom in the function $T_0(B)$
does not allow for a decreasing chiral transition temperature as a function of 
$B$.
This is in agreement with previous mean-field calculations.
}

\begin{document}

\maketitle


\section{Introduction}
\label{sec:introduction}

In recent years QCD in a strong magnetic field has received considerable 
attention. 
This interest has partly been spurred
by non-central heavy-ion collisions at the Relativistic Heavy-Ion Collider
and the Large Hadron Collider. In these experiments, time-dependent
magnetic fields on the order of $|eB|\sim5m_{\pi}^2$ are 
created~\cite{harmen,tonev,bzdak}
and so detailed knowledge of strongly interacting matter in 
external fields is necessary.

At $T=0$, the response of the QCD vacuum to an external magnetic field
is well-known. Lattice calculations as well as calculations
using the Nambu-Jona-Lasinio (NJL) 
model~\cite{lemmer,ebert22,dcsm1,dcsm2,dcsm3,dcsm4}, 
the quark-meson (QM) model~\cite{dcsm6},
Schwinger-Dyson equations for QED~\cite{dcsm5} and QCD~\cite{dcsm7}, 
and the Walecka model~\cite{andreas}
show that the chiral condensate increases
as a function of the external magnetic field $B$.
Moreover, even the weakest magnetic fields induce a chiral condensate
and thus dynamical chiral symmetry breaking if chiral symmetry is intact
at $B=0$. 

The fact that the chiral condensate at zero temperature grows as a function of 
the magnetic field
might lead to the expectation that the critical temperature for the
chiral transition ($T_c$) increases as well. Indeed, mean-field
calculations employing the NJL model or the Polyakov loop extended
NJL (PNJL)~\cite{pnjlgat,pnjlkas} 
model as well as the (P)QM 
model~\cite{fraga1,fragapol,duarte,grecoleik,grecoleik2,rashid} 
show that the critical temperature
is an increasing function of the magnetic field. 
This qualitative behavior is independent of the masses of the $\sigma$
and $\pi$ mesons.
Additionally, the inclusion of mesonic
fluctuations by applying the functional renormalization group (FRG) does not 
qualitatively change
this picture~\cite{skokov,anders,william,kamikado}.

Results from lattice calculations tell a different story. It is seen that the 
behaviour of $T_c$ with $B$ is only increasing at unrealistically large values 
of the pion masses~\cite{sanf,negro}, 
however with physical pion masses $T_c$ is seen to decrease 
with $B$~\cite{buda,budaleik,gunnar,bruckmann,analysis}.
A number of groups have begun altering the standard treatment of chiral models
to include a mechanism for this inverse magnetic catalysis around 
$T_c$~\cite{ferro13,pintoleik,ferro,ferrer14,mintz}.
Two such alterations to the PQM model are to allow either the Yukawa coupling, or the 
transition
temperature of the gluonic sector, to vary with magnetic field.
The former is further motivated by two recent papers utilizing the NJL model 
that were able to
demonstrate inverse magnetic catalysis around $T_c$ by varying the
four-point coupling~\cite{pintoleik,ferro}. However, in a recent paper by Fraga 
et al.~\cite{mintz} it was shown that neither of these freedoms
were sufficient to obtain inverse magnetic catalysis around $T_c$, other than 
for a limited range of low values of the magnetic field.

Motivated both by the seemingly conflicting results coming from the NJL model, 
and the extension of the work of Fraga et al.\ to a 
functional renormalization group (FRG) treatment we 
investigate the effects of varying the Yukawa coupling, $g$,
within the QM model. We use two different regularization schemes, namely
dimensional regularization (DR) and a sharp cutoff. It is seen that simply varying 
$g$
whilst employing a sharp cutoff gives results that are quantitatively and 
qualitatively dependent upon the
scale of the cutoff, whilst using DR one obtains results
that are independent of the renormalization scale. We also investigate varying 
the transition temperature of the gluonic potential within the PQM model using 
the FRG and find, in agreement with the 
prediction of~\cite{mintz}, that the FRG does not allow for inverse magnetic 
catalysis over an extended range of magnetic field values.

The paper is organized as follows. In Sec.~\ref{sec:QM}, we briefly discuss the
quark-meson model. In Sec.~\ref{sec:MF}, we calculate the effective potential
in the mean-field approximation using different regularizations.
In Sec.~\ref{sec:results}, we present our results for the phase diagram
as a function of $g$, $\Lambda_\textrm{UV/DR}$ (the cutoff/DR scale) and finally $T_0$.
In Sec.~\ref{sec:summary} we discuss the results and
briefly summarize our work.


\section{The quark-meson model}
\label{sec:QM}

The quark-meson model is a low-energy effective theory for chiral symmetry in QCD.
In two-flavor
QCD it couples the $O(4)$-symmetric linear sigma model to a massless quark doublet via the 
Yukawa coupling $g$.
The Euclidean Lagrangian in a magnetic field is then given by
\bqa\nonumber
{\cal L}_{}
&=&{1\over2}\left[(\partial_{\mu}\sigma)^2+(\partial_{\mu}\pi_0)^2\right]
+(D_{\mu}\pi^+)^{\dagger}D_{\mu}\pi^++{1\over2}m^2 \big( \varphi^\dagger \varphi \big)
+\frac{\lambda}{24}\big( \varphi^\dagger \varphi \big)^2 -h\sigma
\\ &&
+\bar{\psi}\left[
\gamma_{\mu}D_{\mu}
+g(B)(\sigma-\gamma_5{\boldsymbol\tau}\cdot{\boldsymbol \pi})
\right]\psi\;,
\label{lag}
\eqa
where 
the covariant derivative is $D_{\mu}=\partial_{\mu}-iq_fA_{\mu}^{\rm EM}$,
with $q_f$ a diagonal matrix of the electric charges of the up and down quarks.
${\boldsymbol \tau}$ are the Pauli matrices, $\varphi^\dagger = 
(\sigma,\pi_0,\pi_1,\pi_2)$ and $\pi^{\pm}={1\over\sqrt{2}}(\pi_1\pm i\pi_2)$.
The fermion field is an isospin doublet,
\bqa
\psi=
\left(\begin{array}{c}
u\\
d\\
\end{array}\right)\;,
\label{d0}
\eqa
which, as stated, couples to the mesonic sector via the Yukawa coupling $g(B)$, 
where we have indicated explicitly that this will be allowed to vary with $B$. 
We make
no assumptions as to the manner of this dependence, and simply investigate the
available parameter space when any such dependence is allowed (as was done 
in~\cite{mintz}).

In the absence of external gauge fields the Lagrangian~(\ref{lag})
is $O(4)$ symmetric if $h=0$ and $O(3)$ symmetric if $h\neq0$.
In the presence of a background Abelian gauge field, the $O(4)$ symmetry is 
reduced to an
$O(2)\times O(2)$ symmetry, because of the different electric charges of the
$u$ and $d$ quarks.

Chiral symmetry (or approximate chiral symmetry when $h\neq0$) is broken in the 
vacuum by a nonzero expectation value
$\phi$ for the sigma field. 
Expanding $\sigma$ around this mean field $\phi$ we define 
\bqa
\sigma&=&\phi+\tilde{\sigma}\;,
\eqa
where $\tilde{\sigma}$ is a quantum fluctuating field with vanishing
expectation value. The tree-level potential is then 
\bqa
{\cal V}_0&=&{1\over2}m^2\phi^2+{\lambda\over24}\phi^4-h\phi+{1\over2}B^2\;.
\label{v0}
\eqa


\section{Mean-field approximation}
\label{sec:MF}

In the one-loop approximation, one takes into account the 
Gaussian fluctuations around the mean-field $\phi$.
The one-loop effective potential can then be written as a sum of
the tree-level term~(\ref{v0}) and the one-loop contributions from
the sigma, the pions, and the quarks. Furthermore, it is a
common approximation in the QM model to omit the quantum and thermal
fluctuations of the bosons, i.e.\ treat them at tree level~\cite{dirk,fraga1}.

The one-loop contribution to the effective potential is then given by 
\bqa\nonumber
{\cal V}_1&=&
-\sum_f{\rm Tr}\log\left[
i\gamma_{\mu}D_{\mu}+m_f
\right]
\\
&=& \sum_{P_0,f,n,s}-{|q_fB|\over2\pi}\int_{p_z}\log
\left[P_0^2+p_z^2+m_f^2+|q_fB|(2n+1-s)\right]\;,
\label{trtaken}
\eqa
where the trace is over Dirac and color indices
and in space-time and $m_{u}=m_{d}=g(B)\phi$.
Summing over the Matsubara frequencies in Eq.~(\ref{trtaken}), we find
\bqa\nonumber
{\cal V}_1&=&
-\sum_{f,n,s}{|q_fB|\over2\pi}\int_{p_z}
\left\{
\sqrt{p_z^2+m_f^2+|q_fB|(2n+1-s)}
-2T\log\left[1+e^{-\beta\sqrt{p_z^2+m_f^2+|q_fB|(2n+1-s)}}\right]
\right\}
\;.
\\ &&
\label{sp}
\eqa
The integral over $p_z$ for the zero temperature term
is divergent and is typically regularized 
using dimensional regularization in $d=1-2\epsilon$ dimensions.
The sum over Landau levels is then
subsequently regulated using $\zeta$-function regularization.
The resulting expression is then expanded around $\epsilon=0$
and the poles in $\epsilon$ are removed by wavefunction renormalization
of the gauge field as well as renormalization of the parameters
in the Lagrangian in the usual way. The details of this calculation
can be found in~\cite{rashid}.
The result for the renormalized one-loop effective potential
reads
\bqa\nonumber
{\cal V}_\textrm{DR}&=&{1\over2}m^2\phi^2
+{\lambda\over24}\phi^4 -h\phi
+{N_cm_q^4\over(4\pi)^2}\sum_f\left[
\log{\Lambda_\textrm{DR}^2\over|2q_fB|}+1
\right]
-{N_c\over2\pi^2}\sum_f(q_fB)^2\left[\zeta^{(1,0)}(-1,x_f)\right. \\
&&
\left.+{1\over2}x_f\log{x_f}\right]
-N_c\sum_{s,f,k}{|q_fB|T\over\pi^2}\int_0^{\infty}dp
\log\left[1+e^{-\beta\sqrt{p^2+M_q^2}}\right]\;,
\label{vdimreg}
\eqa
where $x_f={m_f^2\over|2q_fB|}$ and $M_f^2=\sqrt{m_f^2+|q_fB|(2k+1-s)}$, $\Lambda_\textrm{DR}$ is the renormalization scale
associated with the modified minimal subtraction scheme and
$\zeta(a,x)$ is the Hurwitz zeta-function.

Although not often employed
in renormalizable theories, there is nothing that prevents from using a sharp 
ultraviolet cutoff $\Lambda_\textrm{UV}$ to regulate the divergent integrals.
The effective potential is then

\bqa\nonumber
{\cal V}_\textrm{cut}&=&{1\over2}m^2\phi^2+{\lambda\over24}\phi^4-h\phi
+{2N_c\over(4\pi)^2}\sum_f
\bigg\{-\Lambda_\textrm{UV}\sqrt{\Lambda_\textrm{UV}^2+m_f^2}(2\Lambda_\textrm{UV}^2+m_f^2) 
\\ && \nonumber
+m_f^4\log{{\Lambda_\textrm{UV}+\sqrt{\Lambda_\textrm{UV}^2+m_f^2}\over m_f}}
- {1\over4}m_f^4
-4(q_fB)^2\left[\zeta^{(1,0)}(-1,x_f)-{1\over2}(x_f^2-x_f)\log x_f
\right]
\bigg\}
\\ &&
-N_c\sum_{s,f,k}{|q_fB|T\over\pi^2}\int_0^{\infty}dp
\log\left[1+e^{-\beta\sqrt{p^2+M_q^2}}
\right]
\;.
\label{vcutoff}
\eqa
Comparing the two expressions, Eqs.~(\ref{vdimreg}) and (\ref{vcutoff}), we see 
that they have similar structure, other than the presence of an additional term 
coupling various powers of $m_{f}=g\phi$ and $\Lambda_\textrm{UV}$. Additionally the finite 
temperature term is independent of the regularization scheme.

The effective potential ${\cal V}_\textrm{DR}(\phi)$ (${\cal V}_\textrm{cut}(\phi)$) depends upon the parameters 
$\lambda$, $m^2$, $g$, $h$ and $\Lambda_\textrm{DR}$ ($\Lambda_\textrm{UV}$). As we will explore the dependence of 
the transition temperatures on $\Lambda_\textrm{DR}$ and $\Lambda_\textrm{UV}$,  these are left as completely `free'.
Without explicit symmetry breaking the pions are true Goldstone
bosons, i.e.\ we need not fix $m_{\pi}$ as it is automatically zero. At nonzero 
pion mass we adjust $h$ to set $m_\pi$. 
The pion decay constant, $f_\pi$, and sigma mass $m_\sigma$ in the vacuum, 
$T=B=0$, are
fixed by tuning $\lambda$ and $m^2$. All these must be tuned for every different
value of $\Lambda_\textrm{DR}$ or $\Lambda_\textrm{UV}$ and of course they will be different for the different regularization 
schemes. We use the values $f_{\pi}=93$ MeV and $m_{\sigma}=530$ MeV throughout 
this paper, and at the physical point $m_\pi=139$ MeV. Finally the Yukawa 
coupling, $g=g(B)$, at $B=0$ is set to 3.2258 such that the constituent quark 
mass is $g\phi=300$~MeV. However at finite $B$ and $T$ we will allow this to 
vary whilst holding $m^2$ and $\lambda$ fixed.


\section{Numerical results}		
\label{sec:results}

Our initial motivation was to use the FRG
to see if the conclusions in Ref.~\cite{mintz} would be altered when
including mesonic fluctuations. It turns out that FRG seems to open for the possibility of 
inverse magnetic catalysis. However, the 
flow equation involves integration of momenta $k$
from a sharp ultraviolet cutoff $\Lambda_\textrm{FRG}$
down to $k=0$ and the naive use of this cutoff proves problematic.

\begin{figure}[htb]
\begin{center}
\setlength{\unitlength}{1mm}
\includegraphics[width=17.0cm]{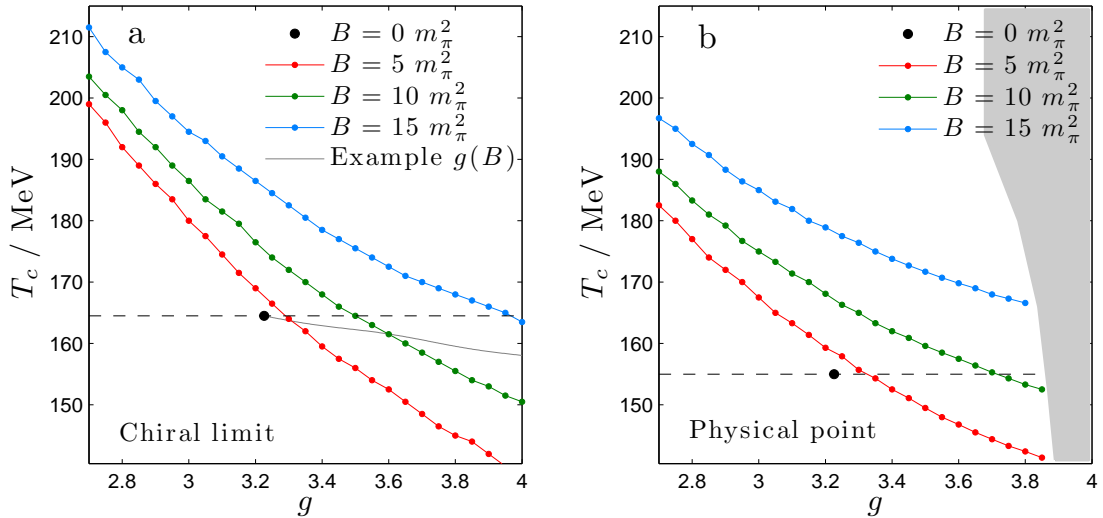}
\caption{Chiral transition temperature as a function of possible value of the Yukawa coupling for the chiral limit (\ref{dimbasic}a) and the physical point (\ref{dimbasic}b). The plots show that a function $g(B)$ starting at $g=3.2258$, $\color{black}{\bullet}$, can give inverse magnetic catalysis up to around $10$ $m_\pi^2$ at the physical point. Beyond this the theory breaks down (grey region). See text for details.}
\label{dimbasic}
\end{center}
\end{figure}

Following Ref.~\cite{mintz}, in Fig.~\ref{dimbasic}, we plot the critical temperature for the 
chiral transition as a function of the Yukawa coupling $g(B)$ for various values of $B$. Figure~\ref{dimbasic}a gives the results in the chiral limit, while Fig.~\ref{dimbasic}b is at the physical point.  The curves are obtained using the dimensionally regulated mean-field result, Eq.~(\ref{vdimreg}), with $\Lambda_\textrm{DR}=182$ MeV, as was used in~\cite{rashid}. The results shown in Fig.~\ref{dimbasic} are in approximate agreement with those of Fraga et al.~\cite{mintz}. At the physical point (\ref{dimbasic}b) we see that the critical temperature becomes undefined at high $B$ and $g(B)$, as given by the grey region. We return to this point shortly.

Figure~\ref{dimbasic} is understood as follows: At $B=0$ the constituent quark 
mass fixes the Yukawa coupling, and thus the chiral transition temperature is 
fixed to be 165 MeV (155 MeV at the physical point), as is given by 
$\color{black}{\bullet}$. The dashed black line is simply a visual guide to 
distinguish catalysis from inverse catalysis. Moving to finite magnetic field 
the value of the Yukawa coupling as a function of $B$ and $T$
is not known, thus we allow for any possible dependence. Any particular function
$g(B)$ is a curve beginning at $\color{black}{\bullet}$ and successively intersecting
the various contours of increasing $B$. One such function $g(B)$ is given by
the grey line in Fig.~\ref{dimbasic}a. If the functional dependence is given by
\begin{equation}
g(B) = g(0)\left[ 1+a(B/m_\pi^2)^b \right]
\end{equation}
then this grey line corresponds approximately to $a=0.0008$ and $b=2.2$.
Here $g$ increases with $B$ in such
a way as to give inverse magnetic catalysis up to at least 12 $m_\pi^2$.
An even simpler function is a 
line moving vertically upwards from $\color{black}{\bullet}$. This corresponds to 
the standard case, where one assumes the Yukawa coupling is independent of $B$ 
and $T$ i.e.\ $a=b=0$. In this case, of course, we find $T_c$, increasing with $B$, i.e.\ magnetic catalysis.

\begin{figure}[htb]
\begin{center}
\setlength{\unitlength}{1mm}
\includegraphics[width=17.0cm]{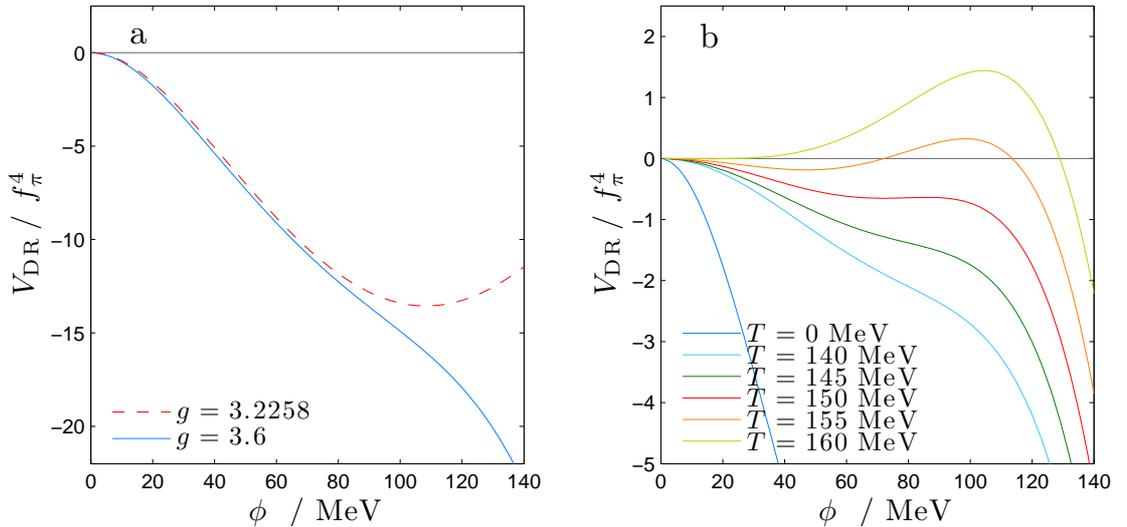}
\caption{\ref{pots}a gives the mean-field potential in the chiral limit for $g=3.2258$ and 3.6 at $T=0$. We see at $g=3.6$ that the theory has an unstable vacuum. \ref{pots}b is the potential with $g=3.6$ for increasing $T$. It is seen that the potential develops a local minimum at positive, finite $\phi$ before the transition temperature, which lies at 161.5 MeV. In both plots $B=10$ $m_\pi^2$.}
\label{pots}
\end{center}
\end{figure}

After inspection of Fig.~\ref{dimbasic}a it seems
quite possible to create a function $g(B)$ such that we have inverse magnetic catalysis over a
large range of magnetic field strength. 
However the complete picture is more complex as is shown at the 
physical point, given in Fig.~\ref{dimbasic}b. Firstly, the change in the 
definition of the critical temperature (from a second order transition, to a 
cross over with a pseudocritical temperature) flattens the curves of 
constant $B$ such that a greater change in $g$ is required for the same change in $B$ and the total range of $B$ values over which one could have inverse catalysis is reduced (for example the blue $B=15$ $m_\pi^2$ curve may never cross the dashed line even if it could be continued to infinitely high $g$). More problematic than this, at large $g$ (given by the grey region), the theory breaks
down. We now explain this with the help of Fig.~\ref{pots}.

As is well known, at very large values of the field $\phi$ the mean-field 
potential becomes unbounded from below, due to the log term in the zero 
temperature expression (Eq.~(\ref{vdimreg}) for the DR scheme). This term is proportional to the fourth power of the quark mass, so it 
is greatly influenced by varying the Yukawa coupling. This is evident from 
Fig.~\ref{pots}a, where we see at $T=0$~MeV, $B=10$~$m_\pi^2$ that changing $g$
from 3.2258 to 3.6, the local minimum disappears altogether giving us unbounded (unphysical) results, indeed for $g>3.35$ 
this is the case.
Thus if $g$ is \textit{only} a function of $B$ then we may not vary it higher than 3.35.
However, in Fig.~\ref{pots}b we now show the change in the potential 
with $T$ with $B=10$ $m_\pi^2$, $g=3.6$ and in the chiral limit. As $T$ 
increases, we first develop a local minimum, like we have in the zero $T$, 
$g=3.2258$ case, and then the potential undergoes the usual chiral phase transition. Because the 
transition temperature in the chiral limit essentially involves investigating the potential around 
$\phi=0$ we are able to define the transition temperature even for very large 
values of $g$ and $B$. But as $g$ is pushed higher and higher, the region where 
we have a local minimum becomes smaller, both in $T$ and $\phi$.
For this reason
at finite $B$ we will allow $g$ to be a function of both $B$ and $T$ such that 
we have the maximum flexibility in $g$ with which to obtain inverse magnetic
catalysis. At the physical point the definition of the pseudocritical temperature involves 
investigation of a finite value of the field $\phi$, in our case where it is 
equal to half the zero temperature value (but note this is not changed when 
investigating the inflection point). In this case even moderate values of $g$ 
and $B$ give unphysical results, as is given by the grey region in
Fig.~\ref{dimbasic}b. This is the primary reason that disallows inverse magnetic
catalysis even allowing $g$ to be a function of both $B$ and $T$.

\begin{figure}[htb]
\begin{center}
\setlength{\unitlength}{1mm}
\includegraphics[width=10.0cm]{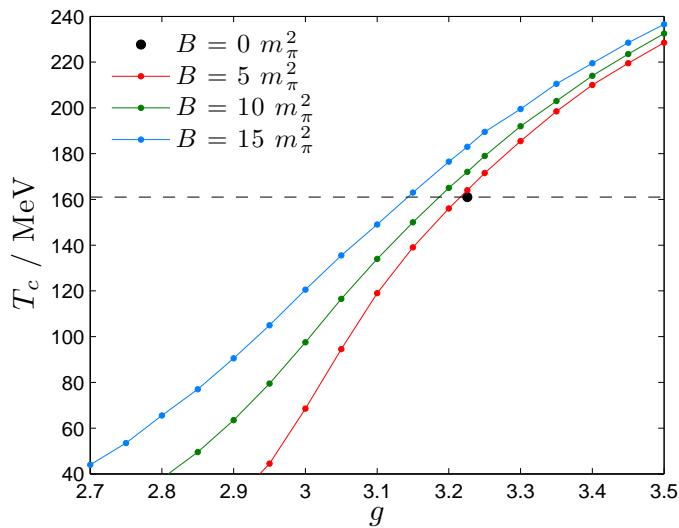}
\caption{$T_c$ plotted against $g$ using a sharp cutoff of 800 MeV. It is seen that a decreasing function $g(B)$ could give inverse magnatic catalysis, however, as noted in the text, this result depends heavily upon the value of the cutoff.}
\label{cutbasic}
\end{center}
\end{figure}

We now turn out attention to the QM model with a sharp cutoff, choosing a value 
of 800 MeV, as was used in our previous calculations using FRG~\cite{anders,william}.
The plot corresponding to the previous Fig.~\ref{dimbasic} is given in 
Fig.~\ref{cutbasic}. These results are qualitatively the same as those we found 
using the FRG. The observed behaviour is reversed as compared with the 
dimensionally regulated result, most obviously $T_c$ is an increasing function 
of $g$ for any fixed $B$ is, while it was a decreasing function in the 
dimensionally regulated theory. In addition we see that it is possible to choose
a function $g(B,T)$ that gives inverse magnetic catalysis. Both as the curves 
become steeper as we move from 3.2258 to approximately 3 (the region of interest
here), but also because $g$ must decrease to obtain inverse magnetic catalysis, 
thus avoiding the problems of an unbounded potential seen above. However we stress
that this result is heavily dependent upon the cutoff used and thus this conclusion 
should not be used out of context. We now turn our attention to this cutoff 
dependence.

\begin{figure}[htb]
\begin{center}
\setlength{\unitlength}{1mm}
\includegraphics[width=10.0cm]{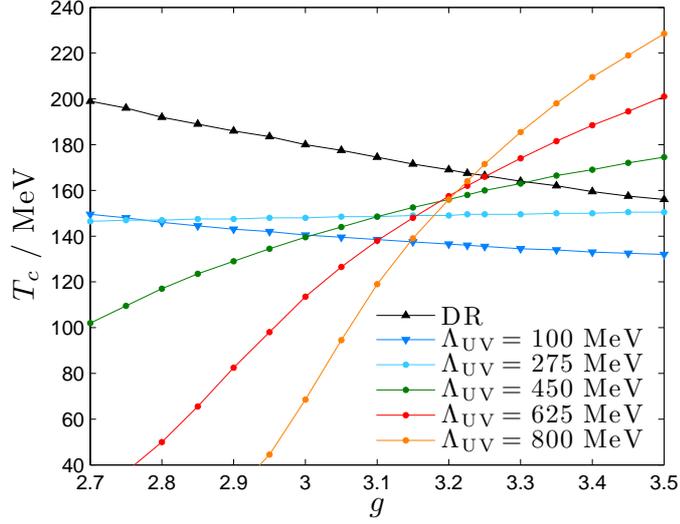}
\caption{$T_c$ with $g(B)$ for various values of $U_\textrm{DR}$ and $U_\textrm{UV}$ for both regularization schemes. For the dimensionally regularized theory only a single curve (black `up' triangles) is given as all values of the renormalization scale between 100 and 800 MeV have $T_c$ within $\pm$ 2 MeV of the curve given, which is calculated using $\Lambda=182$ MeV. In the sharp cutoff theory with $\Lambda=100$ MeV (blue `down' triangles) the phase transition is of first order, all other transition are second order. The plot shows the heavy dependence of the cutoff theory on the value of the cutoff.}
\label{varycut}
\end{center}
\end{figure}

In Fig.~\ref{varycut} we plot $T_c$ for only a single value of the magnetic field, instead
varying the cutoff. The parameter fixing we discussed in Sec.~\ref{sec:MF} is 
done for each different value of $\Lambda_\textrm{DR}$ and $\Lambda_\textrm{UV}$, thus at $g=3.2258$, 
$f_\pi$, $m_{\sigma}$, $m_\pi$ and $m_q$ are equal for each curve. We plot only a 
single value of $\Lambda_\textrm{DR}$ for the theory using DR 
as the results are within $\pm 2$ MeV for all cutoff values between 100 and 800 
MeV. For both regularization schemes the finite $T$ terms are exactly the 
same, and the zero $T$ components are also very similar with exception that the 
sharp cutoff theory adds a term of the form 
$-\Lambda_\textrm{UV}\sqrt{\Lambda_\textrm{UV}^2+m_f^2}(2\Lambda_\textrm{UV}^2+m_f^2)$. Fig.~\ref{varycut} shows that
using a large cutoff this term begins to dominate the behaviour as we increase 
$g$, lowering the potential and thus increasing $T_c$.

\begin{figure}[htb]
\begin{center}
\setlength{\unitlength}{1mm}
\includegraphics[width=10.0cm]{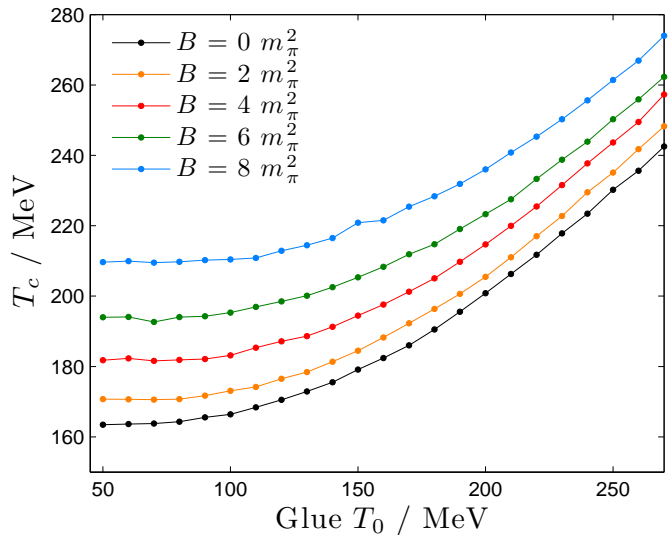}
\caption{$T_c$ with varying $T_0$ for various values of $B$. Note that the scaling in $B$ is more gradual than in previous plots. We see that it is not possible to have inverse magnetic catalysis given any possible function $T_0(B)$ if we also require that $T_c(B=0) \sim T_{d}(B=0)$.}
\label{T0fig}
\end{center}
\end{figure}

It has been suggested that the backreaction of the quarks to the gluons plays a 
primary role in inverse magnetic catalysis~\cite{bruckmann}, thus a natural 
first step towards inverse magnetic catalysis in Polyakov loop coupled models 
would be via tuning the gluonic potential. In \cite{mintz} this was done at 
mean-field level, where they concluded that it was not possible to obtain 
inverse magnetic catalysis by simply varying the gluonic transition 
temperature, $T_0$. We show that this result remains unchanged with the 
inclusion of mesonic fluctuations in Fig.~\ref{T0fig}, which is calculated using
the FRG. We do not introduce the full machinery of the FRG, instead referring 
the reader to~\cite{william}.
The methods are the same as in that paper, other than that we rerun the 
calculation for varying values of $T_0$, which is a free parameter in the model.
For the physical observables $f_\pi$, $m_{f}$, $m_{\pi}$ and $m_{\sigma}$, the 
values
are the same as we used in the physical point, mean-field calculations and 
we have $\Lambda_\textrm{FRG}=800$ MeV. The figure can be read in the same way as 
Fig.~\ref{dimbasic}, but with $T_0$ in the place of $g$. At $B=0$, $T_0$ is 
usually taken as $\sim210$ MeV with two flavours of quarks. This value also 
ensures that $T_c$ is approximately equal to the deconfinement transition 
temperature, $T_d$. Using this as the zero-$B$ starting point we see that the curves 
are simply too flat (and become increasingly so as we decrease $T_0$) to allow 
for inverse magnetic catalysis up to magnetic fields over $B\sim4$ $m_\pi^2$. 
Thus we find in agreement with the prediction of \cite{mintz} that using the 
FRG, there is no possible function $T_0(B)$ which could give inverse magnetic 
catalysis.

\section{Discussion and conclusion}
\label{sec:summary}

In the quark-meson model, one chooses a regularization prescription and fixes the associated scale ($\Lambda=\Lambda_\textrm{DR}$, $\Lambda_\textrm{UV}$, $\Lambda_\textrm{FRG}$) and then sets 
$m^2$, $\lambda$, $h$ and $g$ to obtain the correct values of the particle 
masses and pion decay constant in the vacuum. In doing so the model's 
dependence upon $\Lambda$ is essentially cancelled out. The model is useful 
because we may then vary external parameters without introducing strong 
$\Lambda$ dependence and thus investigate physics outside of the vacuum. But in 
varying $g$ there is no guarantee that the model will be $\Lambda$ independent, 
and as Fig.~\ref{varycut} shows, when using a cutoff this is not the case.
The real problem here, in terms of modelling inverse magnetic catalysis, is not 
only that it is not possible to generate meaningful results at mean-field using a sharp 
cutoff, but that it is not possible to do so using the FRG, as it suffers from exactly the 
same problems.

Usually it is the case the inclusion of mesonic fluctuations has some 
quantitative effect upon the chiral transition temperature but that the results 
remain qualitatively the same. However we have seen from lattice results that 
inverse magnetic catalysis is dependent upon the pion mass. But usually in model calculations
only the mesonic fluctuations are dependent upon the pion mass, thus indicating their importance.
Moreover varying the pion mass amounts to 
varying $m^2$ and $\lambda$ in Eq.~(\ref{vdimreg}) or~(\ref{vcutoff}). But as 
there 
is no coupling between $g$ and either of these variables, varying $m_\pi$ simply 
shifts all of the curves in Fig.~\ref{dimbasic} either up or down, something we 
have checked explicitly. Thus to fully reproduce the lattice results at mean-field by 
varying $g$ (if it was even possible) one would need $g(B,T,m_\pi)$.

We agree with the basic result of Fraga et al.~\cite{mintz}, that within the QM 
model it is not possible to reproduce lattice results by simply utilizing 
$g(B)$, even if we allow complete freedom in this functional dependence. 
However, as we use different regularization the reason for this is very 
different. Moreover we find within our own results that the transition 
temperature as a function of $g$ depends in detail of how one treats the vacuum 
fluctuations.
It is not only a question of whether to include them or not,
as in the case of the order of the transition, but it also depends upon
the exact implementation of the regularization scheme. Allowing $g$ to run
with $B$ acknowledges that there exists physics not captured by the quark-meson
model yet vital in mapping out the chiral phase diagram. But this physics must
be incorperated in such a way that reliable computations can still be made.
This is not the case when simply varying $g$ using a sharp cutoff.





\end{document}